# A Generalized Mathematical Framework for Thermal Oxidation Kinetics

Zhijie Xu[1, a)], Kevin M. Rosso[2] and Stephen M. Bruemmer[3]

1. Computational Mathematics Group, Fundamental and Computational Sciences Directorate, Pacific Northwest National Laboratory, Richland, WA 99352, USA

2. Chemical and Materials Sciences Division, Fundamental and Computational Sciences Directorate, Pacific Northwest National Laboratory, Richland, WA 99352, USA

3. Energy and Environment Directorate, Pacific Northwest National Laboratory, Richland, WA 99352, USA

We present a generalized mathematical model for thermal oxidation and the growth kinetics of oxide films. The model expands long-standing classical models by taking into account the reaction occurring at the interface as well as transport processes in greater detail. The standard Deal-Grove model (the linear-parabolic rate law) relies on the assumption of quasi-static diffusion that results in a linear concentration profile of, for example, oxidant species in the oxide layer. By relaxing this assumption and resolving the entire problem, three regimes can be clearly identified corresponding to different stages of oxidation. Namely, the oxidation starts with the reaction-controlled regime (described by a linear rate law), is followed by a transitional regime (described by a logarithmic or power law depending on the stoichiometry coefficient *m*), and ends with the well-known diffusion-controlled regime (described by a parabolic rate law). Deal-Grove's theory is shown to be the lower order approximation of the proposed model. Various oxidation rate laws are unified into a single model to describe the entire oxidation process.

**Key words:** model, oxidation, linear, logarithm, parabolic, Deal-Grove

[a)] Electronic mail: zhijie.xu@pnl.gov

## I. Introduction

The purpose of this paper is to present a generalized mathematical framework for understanding the growth of oxide films by transport processes taking place in the oxide layer, and oxidation reactions at the oxide-material interface. The term “material’ here represents any underlying materials (metals or metalloids) under consideration. Such transport processes and oxidation reactions at the interface have important roles in forming oxide films, i.e., a “corrosion” process accompanied with undesirable mass loss of the material that is mathematically similar to the “solute precipitation” process.[1,2] A layer of oxide film can be formed naturally under certain conditions and provide the underlying material protection against further corrosion.

Fundamental description of oxidation was laid down in early 1900[3] by Tammann and Pilling and Bedworth.[4] They established the classical parabolic oxidation rate law that modern oxidation theory is based upon.[5,6,7] In classical oxidation theory, the oxidation mechanism considered the diffusion of a chemical species (mainly oxygen) through the oxide layer as the rate-limiting process. The kinetics of oxide formation under the diffusion-controlled condition was later described by Rhines,[8] Darken,[9] and Wagner[10,11] that leads to a parabolic rate law. This simple model, though not able to predict oxidation behavior in all practical applications, nonetheless remains very useful for understanding its most important features and gaining essential knowledge for more complex systems.

Thermal oxidation refers to a process to produce a thin layer of oxide on the surface of given material. For example, it is the process used to generate a thin silicon dioxide layer on the surface of silicon, with numerous applications in the silicon-dominated semiconductor industry. The associated chemical reactions are $Si + O_2 \rightarrow SiO_2$ for dry oxidation and $Si + 2H_2O \rightarrow SiO_2 + 2H_2(g)$ for wet oxidation, respectively. In the vision of the original Deal-Grove's model,[12] thermal oxidation involves three critical steps in accordance with the experimental evidence showing that oxidation proceeds through the continuous inward diffusion of oxidant species and reactions at the silicon dioxide-silicon interface.[13,14,15] The three steps are

i. Transport and dissolution of the oxidizing agent (for example $O_2$) at the external surface;

ii. Diffusion of the oxidizing agent through the oxide towards the oxide-material interface;

iii. Chemical reaction of the oxidizing agent with the reactive element at the oxide-material interface.

Steps ii and iii are demonstrated in Fig. 1a. Continuous production of oxide from the reaction leads to a moving interface at a velocity of $V_s$. It is well known that for many oxidation processes, particularly for thick oxide films, the rate of oxide growth follows the parabolic law, $X_o^2 = 2\gamma_o t$, where $X_o$ is the thickness of the oxide, $\gamma_o$ is the parabolic constant, and $t$ is the oxidation time. This is characteristic of a diffusion-controlled process where the oxidant must travel an increasingly longer distance to the oxide-material interface with increasing oxide thickness. On the other hand, at the earliest stage of oxidation the process can be modeled by a simple linear law, $X_o = C\left(t + \tau\right)$, where $C$ is linear constant and time $\tau$ accounts for the formation of the initial oxide layer at the beginning of oxidation. The linear rate law is often valid for reaction-controlled oxidation processes.

A combined linear-parabolic model was first developed by Deal and Grove[12] in 1965, around the same time when silicon semiconductor technology began. In the original Deal-Grove model, a linear oxidant concentration profile (dashed line in Fig. 1a) was assumed as the result of the quasi-static diffusion assumption. By assuming a first order reaction at the interface, the flux corresponding to the oxidation is expressed as $F_o = k\, C_o\big|^-$, standing for the consumption of oxidant at the interface, where $k$ is the reaction rate constant and $C_o\big|^-$ is the interface concentration of oxidant at the oxide side. The diffusion flux can be written as $F_D = D_o\left(C_o^\infty - C_o\big|^-\right)\big/X_o$ based on the linear concentration profile, where $D_o$ is the diffusion coefficient of the oxidant, and $C_o^\infty$ is the concentration of oxidant at gas/oxide interface as shown in Fig. 1a. By considering mass conservation at the oxide-material interface, we have the relationship

$$F_D = F_o = V_s \rho\,, \tag{1}$$

where $V_s = dX_o/dt$ is the oxide-material interface movement velocity and $\rho$ ($mol/L^3$, $L$ is the unit of length) is the molar density of the oxide. The Deal-Grove model can be obtained from Eq. (1) as

$$\frac{dX_o}{dt} = \frac{k}{1 + kX_o/D_o}\frac{C_o^\infty}{\rho}\,. \tag{2}$$

The solution to differential Eq. (2) leads to the linear-parabolic growth law.

The Deal-Grove model has been widely accepted since 1965, and has been shown to be accurate over a wide range of temperatures, oxide thickness, and oxidant partial pressures. Despite its success, the validity of the Deal-Grove model has been a subject of continual discussion.[16,17,18,19] In particular, there are modeling[17] and experimental studies[20] suggesting that a logarithmic rate law more generally provides a better description for thermal oxidation, which led to the reexamination of the Deal-Grove model presented in this paper. By relaxing the quasi-static diffusion assumption in the original Deal-Grove model, the new model naturally leads to the logarithmic (or power depending on the stoichiometry coefficient *m*) law at the early oxidation stage, followed by the parabolic law at long oxidation time.

The current model was developed for oxide growth due to reaction at and movement of the material-oxide interface. However, conceptually, the same model can be extended to oxide growth at the gas-oxide interface (as described in Fig. 1b), if the physical processes and their governing equations are the same as the growth at the material-oxide interface.

## II. Generalized mathematical framework for thermal oxidation

The generalized mathematical model provides the governing equations of the moving interface problem for thermal oxidation. The simplest thermal oxidation model includes the diffusion of oxidant species in the oxide, and the chemical reactions at the oxide-material interface. The dynamics of the moving interface during oxidation is a result of the competition between the transport of oxidant to the interface and the consumption of the oxidant due to the oxidation reaction at the interface. The system of equations for thermal oxidation first includes the transport equations for oxidant species:

$$\partial C_o / \partial t = D_o \nabla^2 C_o , \tag{3}$$

where $C_o(x,t)$ is the oxidant concentration at position *x* and time *t*, and $\nabla$ is the Laplace operator. The flux of oxidant agent into the oxide-material interface should balance the consumption due to the oxidation reaction. By assuming the oxidation reaction at the oxide-material interface in the form of $M + mO_2 \rightarrow MO_{2m}$, the rate equation can be written as:

$$\frac{d\left(C_{MO_{2m}}\right)}{dt} = k\left(C_o\big|^{-}\right)^m , \tag{4}$$

where $k$ is the interface reaction rate constant and $C_{MO_{2m}}$ is the interface concentration of the oxidation product. Therefore,

$$V_s = -\frac{D_o\left(\nabla C_o\right|^-\right)\cdot\vec{n}}{m\rho} = \frac{k\left(C_o\right|^-\right)^m}{\rho} \text{ at the interface } \Gamma. \tag{5}$$

$C_o|^-$ and $\nabla C_o|^-$ are the oxidant concentration and concentration gradient at the interface with $|^-$ indicating the magnitude of a variable at the oxide side of the interface, $\vec{\mathbf{n}}$ is the unit vector perpendicular to the interface , $\Gamma$, pointing into the material side. Equation (5) defines the interface movement velocity and is a result of the local mass conservation condition.[1]

The above system of equations can be rewritten in dimensionless form by introducing the unit of length $L$, unit of time $S = L^2/D_o$, unit of velocity $U = D_o/L$, and dimensionless number $D_a = kL\rho^{m-1}/D_o$ that represents the ratio between reaction and diffusion rates. The new equations read:

$$\partial c_o / \partial t = \nabla^2 c_o, \tag{6}$$

and

$$v_s = -\frac{1}{m}\left(\nabla c_o\right|^-\right)\cdot\vec{n} = D_a\left(c_o\right|^-\right)^m \text{ on } \Gamma, \tag{7}$$

where the concentration is normalized by $\rho$ ($c_o = C_o/\rho$), the molar density of the oxide product. The solution to Eqs. (6)-(7) is only dependent on the dimensionless number $D_a$, stoichiometry coefficient $m$, and the relevant boundary conditions. Dimensionless number $D_a$ represents the different oxidation regimes with $D_a \to \infty$ corresponding to the diffusion-controlled regime and $D_a \to 0$ corresponding to the reaction-controlled regime. $D_a$ is a function of temperature as both reaction rate $k$ and diffusion coefficient $D_o$ are temperature dependent.

## III. Analytical solutions for 1D diffusion-controlled oxidation (infinite $D_a$)

It is useful to first consider the analytical solution available to the 1D diffusion-controlled case, i.e., where the dimensionless ratio between reaction and diffusion rates $D_a \to \infty$. In this case, the original Eqns. (6)-(7) are reduced to

$$\partial c_o / \partial t = \partial^2 c_o \big/ \partial x^2 \, . \qquad (8)$$

The interface oxidant concentration is reduced to

$$c_o\big|^- = 0 \text{ on } \Gamma , \qquad (9)$$

and the interface velocity is

$$v_s = -\frac{1}{m}\left(\nabla c_o\big|^-\right)\cdot\vec{n} \text{ on } \Gamma . \qquad (10)$$

Analytical solution to Eqs. (8)-(10) can be found as

$$X\left(t\right) = 2\gamma\sqrt{t} \, , \qquad (11)$$

$$v_s\left(t\right) = \partial X / \partial t = \gamma \big/ \sqrt{t} \, , \qquad (12)$$

$$c_o = c_o^\infty \left(1 - \frac{erf\left(x/2\sqrt{t}\right)}{erf\left(\gamma\right)}\right), \qquad (13)$$

where $X$ is the non-dimensional oxide thickness. It is clear from this relationship that diffusion-controlled oxidation follows the parabolic rate law. The parabolic constant $\gamma$ can be determined from the relevant boundary conditions using:

$$m\sqrt{\pi}\gamma \cdot erf\left(\gamma\right)\exp\left(\gamma^2\right) = c_o^\infty . \tag{14}$$

It is worth mentioning that for oxidation with sufficiently long duration, the interface oxidant concentration $c_o\big|^-$ keeps decreasing and eventually approaches zero due to the thickening of the oxide layer and consequently longer distance that the oxidant has to travel to the interface. Therefore, oxidation with finite $D_a$ will eventually enter the diffusion-controlled regime and follow the parabolic rate law for sufficiently long oxidation time. The issue addressed by the present study is to determine the oxidation rate for the transitional period before entering the diffusion-controlled regime, and how long it takes for the transition to complete.

## IV. Solutions for 1D oxidation with finite $D_a$

It is not trivial to solve Eqs. (6) and (7) with a finite dimensionless number $D_a$. As depicted in Fig. 2, we first introduce the following relationships between the interface values and the interface velocity through a straightforward differential analysis,

$$\left.\frac{\partial c_o}{\partial t}\right|^- = \frac{\partial \left. c_o\right|^-}{\partial t} - \left.\frac{\partial c_o}{\partial x}\right|^- \cdot v_s , \tag{15}$$

$$\left.\frac{\partial\left(\partial c_o/\partial x\right)}{\partial t}\right|^- = \frac{\partial\left.\left(\partial c_o/\partial x\right)\right|^-}{\partial t} - \left.\frac{\partial^2 c_o}{\partial x^2}\right|^- \cdot v_s . \tag{16}$$

Similarly, other higher order derivatives at the interface can be obtained in the same fashion,

$$\left.\frac{\partial\left(\partial^n c_o/\partial x^n\right)}{\partial t}\right|^- = \frac{\partial\left.\left(\partial^n c_o/\partial x^n\right)\right|^-}{\partial t} - \left.\frac{\partial^{n+1} c_o}{\partial x^{n+1}}\right|^- \cdot v_s , \tag{17}$$

$$\left.\frac{\partial^{n+2}c_o}{\partial x^{n+2}}\right|^{-} = \left.\frac{\partial^n\left(\partial c_o/\partial t\right)}{\partial x^n}\right|^{-} = \left.\frac{\partial\left(\partial^n c_o/\partial x^n\right)}{\partial t}\right|^{-}, \; n = 0,1,2,3...... \tag{18}$$

By substituting the interface values into the interface conditions (Eq. (7)), we arrived at the equations for interfacial concentration and corresponding derivatives for oxidant concentration up to the third order,

$$c_o|^{-} = \left(v_s/D_a\right)^{1/m}, \tag{19}$$

$$\left.\frac{\partial c_o}{\partial x}\right|^{-} = -mv_s, \tag{20}$$

$$\left.\frac{\partial^2 c_o}{\partial x^2}\right|^{-} = \left.\frac{\partial c_o}{\partial t}\right|^{-} = \frac{\partial\, c_o|^{-}}{\partial t} + mv_s^2 = \frac{1}{mv_s}\left(\frac{v_s}{D_a}\right)^{1/m}\frac{\partial v_s}{\partial t} + mv_s^2, \tag{21}$$

$$\left.\frac{\partial^3 c_o}{\partial x^3}\right|^{-} = \left.\frac{\partial\left(\partial c_o/\partial x\right)}{\partial t}\right|^{-} = -m\frac{\partial v_s}{\partial t} - \frac{1}{m}\left(\frac{v_s}{D_a}\right)^{1/m}\frac{\partial v_s}{\partial t} - mv_s^3. \tag{22}$$

In principle, any higher order concentration derivatives ( $\left.\frac{\partial^4 c_o}{\partial x^4}\right|^{-}$, $\left.\frac{\partial^5 c_o}{\partial x^5}\right|^{-}$ ,……) can be obtained in a similar manner. On the other hand, the concentration field can be written in terms of those derivatives through Taylor expansion,

$$c_o^{\infty} = c_o|^{-} + \sum_{n=1}^{\infty}\frac{(-X)^n}{n!}\left.\frac{\partial^n c_o}{\partial x^n}\right|^{-} = c_o|^{-} - X\left.\frac{\partial c_o}{\partial x}\right|^{-} + \frac{X^2}{2!}\left.\frac{\partial^2 c_o}{\partial x^2}\right|^{-} - \frac{X^3}{3!}\left.\frac{\partial^3 c_o}{\partial x^3}\right|^{-} + ...... \tag{23}$$

It should be pointed out that the original Deal-Grove model only includes the first two terms on the RHS side of Eq. (23) due to the quasi-static diffusion assumption. The formulation presented in this paper considers the contribution from higher order terms that are essential for the appearance of transitional regime (logarithmic or power law) between linear and parabolic rate laws in the model. By substituting expressions for interface concentration and derivatives (Eqs. (19)-(22)) into Eq. (23), we have the expression of $c_o^\infty$ in terms of interface velocity $v_s$ and oxide thickness $X$ ,

$$c_o^\infty \approx \left(\frac{v_s}{D_a}\right)^{1/m} + m\left\{\exp\left(v_s X\right) - 1\right\} + \frac{1}{m v_s^3}\left(\frac{v_s}{D_a}\right)^{1/m} \frac{\partial v_s}{\partial t}\left\{\exp\left(v_s X\right) - v_s X - 1\right\}. \tag{24}$$

This can be further rewritten in terms of $v_s$ ,

$$\frac{\partial v_s}{\partial t} = -\lambda \cdot \frac{m^2 v_s^3}{\left(v_s / D_a\right)^{1/m}}, \tag{25}$$

and

$$\lambda = \frac{\left\{\exp\left(v_s X\right) - 1\right\} - \left\{c_o^\infty - \left(v_s / D_a\right)^{1/m}\right\} / m}{\left\{\exp\left(v_s X\right) - 1\right\} - v_s X}, \tag{26}$$

where $X\left(t\right) = \int_0^t v_s\left(t'\right) dt'$ and $v_s = dX/dt$ . Although an analytical solution to Eq. (25) cannot be found, several remarkable findings can be made:

1) At the early oxidation stage we have the following approximation,

$$c_o^\infty \approx \left(v_s / D_a\right)^{1/m} + m v_s X \tag{27}$$

leading to $\lambda \approx 1$ . Eq. (25) can be reduced to

$$\frac{\partial v_s}{\partial t} = -\frac{m^2 v_s^3}{\left(v_s / D_a\right)^{1/m}}. \tag{28}$$

The analytical solution to Eq. (28) can be found as

$$v_s = \gamma_1 t^{-n}, \tag{29}$$

where $n = \dfrac{1}{2 - 1/m}$ and $\gamma_1 = \left(\dfrac{n}{m^2 D_a^{1/m}}\right)^n$. The simplest situation of $m = 1$ leads to $n = 1$ and $\gamma_1 = 1/D_a$, and oxidation follows the logarithmic law, i.e., $X(t) \propto \log(t)$. For the other stoichiometry coefficient *m,* a power law can be expected, i.e., $X(t) \propto t^{1-n}$.

2) At long oxidation duration $t \to \infty$ we have the following approximation

$v_s \approx 0$ and $\partial v_s / \partial t \approx 0$ leading to

$$c_o^\infty \approx m\left\{\exp(v_s X) - 1\right\} \text{ and } \lambda \approx 0 \quad (30)$$

the analytical solution becomes

$$v_s = \gamma_2 t^{-1/2}, \quad (31)$$

where $\gamma_2 = \left(l\,\mathrm{og}\sqrt{1 + c_o^\infty / m}\right)^{1/2}$. The oxidation eventually follows the parabolic rate law, i.e. $X(t) \propto t^{1/2}$.

3) As shown in Fig. 3, a typical oxidation process can be divided into three distinct regimes, i.e., the reaction-controlled linear regime ($0 < t < t_1$), where $v_s \approx D_a \left(c_o^\infty\right)^m$, followed by the transitional regime ($t_1 < t < t_2$) where $v_s \approx \gamma_1 t^{-n}$, and the diffusion-controlled regime ($t > t_2$), where $v_s \approx \gamma_2 t^{-1/2}$. A rough estimation of time $t_1$ and $t_2$ can be made as

$$D_a^2 t_1 = \frac{1}{m(2m-1)\, c_o^{\infty(2m-1)}}, \quad (32)$$

$$D_a^2 t_2 = \frac{1}{\left[m(2m-1)\right]^{2m} \left[\log\sqrt{1 + c_o^\infty / m}\right]^{(2m-1)}}. \quad (33)$$

Obviously, the duration of each regime scales with $1/D_a^2$ and is dependent on $c_o^\infty$ and *m*. Figure 4 presents the variation of time $t_1$ and $t_2$ with $c_o^\infty$ for $m$ = 1. Hence, the observed oxidation rate can follow any one of the three rate laws depending on the oxidation time and the dimensionless number $D_a$ , i.e., the competition between reaction and diffusion processes.

**V. Comparison of Oxidation Models**

Through substituting relationship $v_s = dX/dt$ into Eq. (25), original Eq. (25) can be rewritten into a second order Ordinary Differential Equation (ODE) in terms of *X*.

$$\frac{\partial^2 X}{\partial t^2} = -\frac{\left\{\exp\left(X\,\partial X/\partial t\right)-1\right\}-\left\{c_o^\infty-\left(\partial X/\partial t/D_a\right)^{1/m}\right\}\Big/m}{\left\{\exp\left(X\,\partial X/\partial t\right)-1\right\}-X\,\partial X/\partial t}\cdot\frac{m^2\left(\partial X/\partial t\right)^3}{\left(\partial X/\partial t/D_a\right)^{1/m}} \qquad (34)$$

with initial condition $v_s\big|_{t=0} = D_a\left(c_o^\infty\right)^m$ and $X\big|_{t=0} = 0$ . Equation (34) can be numerically solved with any arbitrary dimensionless number $D_a$ , boundary condition $c_o^\infty$ , and stoichiometry coefficient *m*. For the purpose of comparison, we present the solutions to the Deal-Grove model,

$$v_s = \frac{D_a c_o^\infty}{\left(1+2D_a^2 c_o^\infty t\right)^{1/2}}, \qquad X = \frac{1}{D_a}\left\{\left(1+2D_a^2 c_o^\infty t\right)^{1/2}-1\right\} \qquad (35)$$

and the solutions to the logarithmic model,

$$v_s = \frac{D_a c_o^\infty}{1+D_a^2 c_o^\infty t}, \qquad X = \frac{1}{D_a}\ln\left(1+D_a^2 c_o^\infty t\right) \qquad (36)$$

with the same initial conditions.

Figure 5 presents a comparison between the current model, the Deal-Grove model and the logarithm model for the log-log dependence of interface velocity $v_s$ on oxidation time *t*, with the solid line representing the current model, dashed line representing the Deal-Grove model, and the dotted line for the logarithm model. At the early oxidation stage $0 < t < t_1$, all three models lead to a similar linear regime. The current model and logarithm model agree with each other up to $0 < t < t_2$. All three rate laws observed in a large number of oxidation experiments appear in the current oxidation model representing different stages in the entire oxidation process with the logarithm (or power) rate law representing the transitional between the linear and parabolic rate laws. However, the original Deal-Grove model is limited to only the linear and parabolic regimes without a transitional regime in between.

Figure 6 shows a similar comparison between the current model and the Deal-Grove model on a log-log plot for the interface velocity $v_s$ and oxide thickness *X*. For completeness, the numerical solutions for the variation of $\log_{10} v_s$ on the thickness *X* and the dependence of thickness *X* on the oxidation time *t* are also presented in Fig. 7 and Fig. 8, respectively. Finally, the effect of stoichiometry coefficient *m* on the oxidation kinetics is shown in Fig. 9 with *m* = 1, 2, and 3 for $D_a = 1$ and $c_o^\infty = 1 \times 10^2$.

It was confirmed that the logarithmic growth law is more general than the linear-parabolic law in comparison to the experimental results for the entire oxidation regime for the growth of silicon oxides by dry oxidation.[17] For a relatively thin oxidation layer or a short oxidation time, the fast oxidation is more likely controlled by the combination of both reaction and transport processes (namely oxidation falls in the transition regime), instead of either a pure reaction-controlled or diffusion controlled process. The present model provides the potential theoretical foundation for describing the transitional oxidation behavior by complementing the linear-parabolic rate law. It is also certainly interesting to apply the proposed model to materials other than silicon, where the major oxidation mechanism can be described by the competition between the oxidation reaction and transport processes.

## VI. Conclusion

A generalized mathematical model for thermal oxidation and rigorous solution are presented for any arbitrary dimensionless number $D_a$ (the ratio between reaction and diffusion rate), boundary condition $c_o^\infty$, and stoichiometry coefficient *m*. The typical solution to the proposed model exhibits three oxidation rate laws for the entire oxidation process, with the linear law corresponding to the reaction-controlled regime, the logarithmic or power law corresponding to the transitional regime, and the parabolic law corresponding to the diffusion-controlled regime. The standard Deal-Grove model can only lead to the linear-parabolic rate law and was shown to be the lower order approximation of the presented model. Therefore, the current model offers much greater descriptive range for the entire oxidation process. Ongoing studies include comparison of the current model with numerical modeling and experimental data.

## ACKNOWLEDGMENTS

This research was supported by a grant from the U.S. Department of Energy, Office of Science, Office of Basic Energy Sciences, Materials Science program.

Table 1. Definition and units of symbols and variables.

| Symbol | Units | Definition |
| --- | --- | --- |
| $X_o$ | $L$ | Oxide thickness |
| $\gamma_o$ | $L^2/S$ | Parabolic constant |
| $k$ | $L^{3m-2}/(mol^{m-1} \cdot S)$ | Reaction rate constant |
| $F_o$ | $mol/(L^2 \cdot S)$ | Oxidation reaction flux |
| $F_D$ | $mol/(L^2 \cdot S)$ | Diffusion flux |
| $D_o$ | $L^2/S$ | Oxidant diffusion coefficient |
| $C_o$ | $mol/L^3$ | Oxidant concentration |
| $C_o^\infty$ | $mol/L^3$ | Oxidant concentration at gas-oxide interface |
| $V_s$ | $L/S$ | Interface moving velocity |
| $\rho$ | $mol/L^3$ | Molar density of the oxide |
| $m$ | Dimensionless | Stoichiometry coefficient |
| $c_o$ | Dimensionless | Oxidant concentration |
| $c_o^\infty$ | Dimensionless | Oxidant concentration at gas-oxide interface |
| $v_s$ | Dimensionless | Interface moving velocity |
| $D_a$ | Dimensionless | Ratio between reaction and diffusion |
| $\vec{\mathbf{n}}$ | Dimensionless | Unit vector on oxide-material interface |
| $\gamma$ | Dimensionless | Parabolic constant |
| X | Dimensionless | Oxide thickness |

Figure 1. Schematic plot of thermal oxidation and the oxidant concentration profile. a) oxidation due to the growth of the oxide-material interface and b) oxidation due to the growth of the gas-oxide interface.

Figure. 2. Schematic plot of the moving interface used to derive the differential relationships between interface values and interface velocity (Eqs. (15) and (16)).

Figure. 3. Schematic plot of a typical oxidation process divided into three distinct regimes.

Figure. 4. Variation of time $t_1$ (dashed line) and $t_2$ (solid line) with $c_o^\infty$ for $m$ = 1.

Figure. 5. Log-log plot of the dependence of interface velocity $v_s$ on oxidation time $t$ for different values of $c_o^\infty$ with dimensionless number $D_a = 1$ and $m$ = 1.

Figure. 6. Log-log plot of the dependence of interface velocity $v_s$ on oxide thickness $X$ for different values of $c_o^\infty$ with dimensionless number $D_a = 1$ and $m$ = 1.

Figure. 7. Log plot of the dependence of interface velocity $v_s$ on oxide thickness $X$ for different values of $c_o^\infty$ with dimensionless number $D_a = 1$ and $m$ = 1.

Figure. 8. Plot of dependence of oxide thickness $X$ on oxidation time $t$ for different values of $c_o^\infty$ with dimensionless number $D_a = 1$ and $m$ = 1.

Figure. 9. Log-log plot of the dependence of interface velocity $v_s$ on oxidation time $t$ for m = 1, 2, and 3 with dimensionless number $D_a = 1$ and $c_o^\infty = 1 \times 10^2$.

Fig.1.

(a)

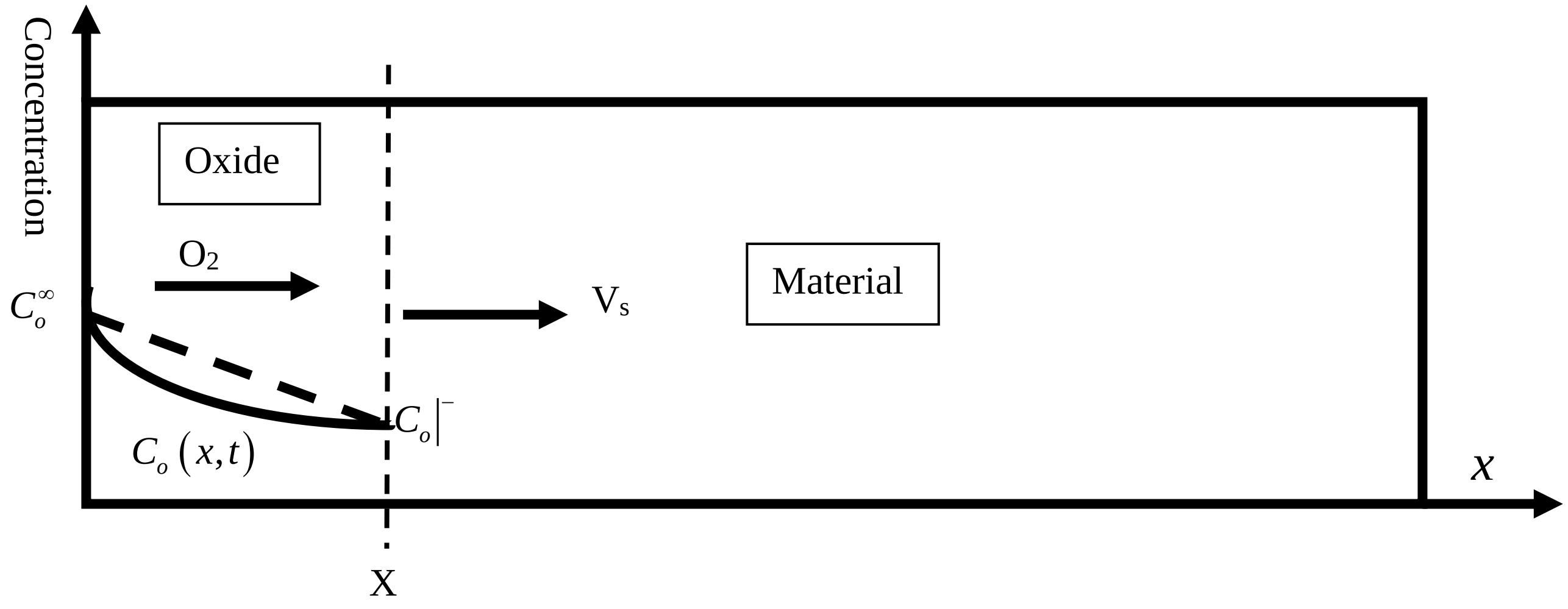

(b)

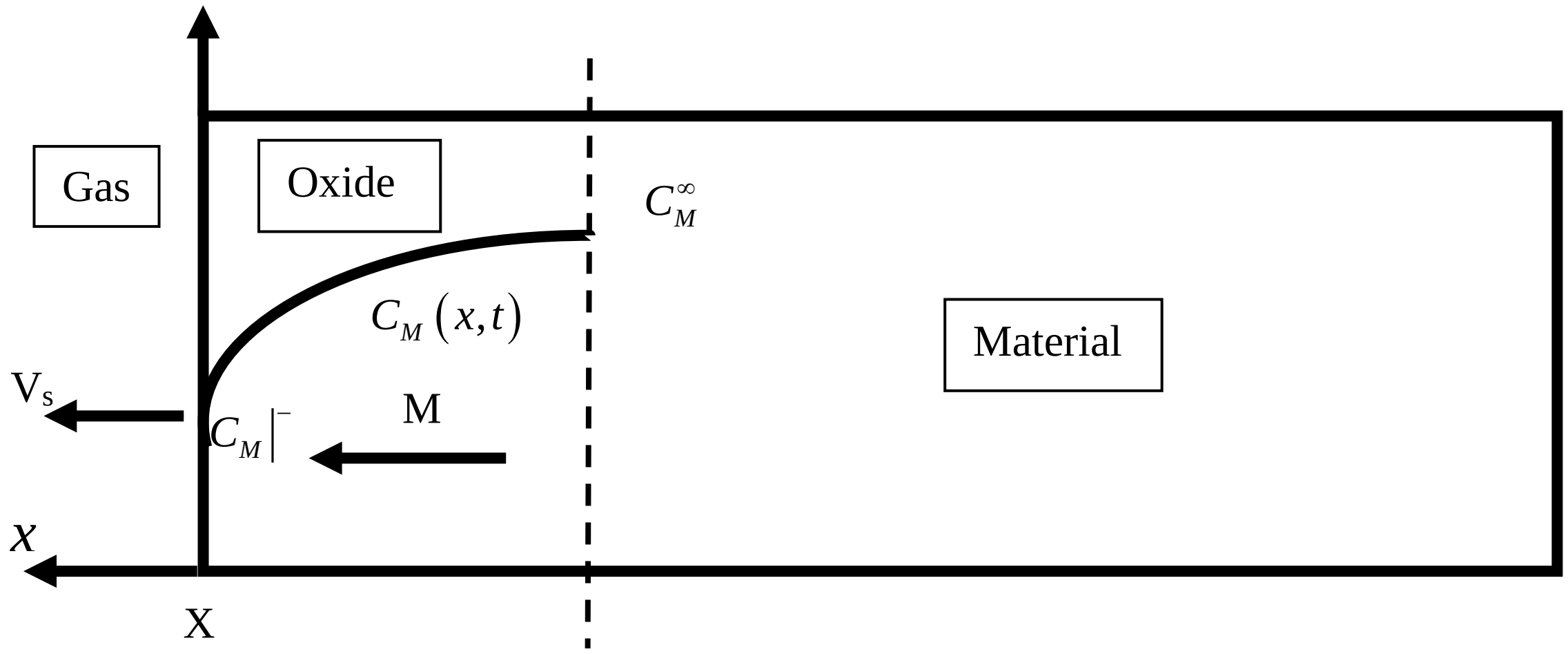

Fig.2.

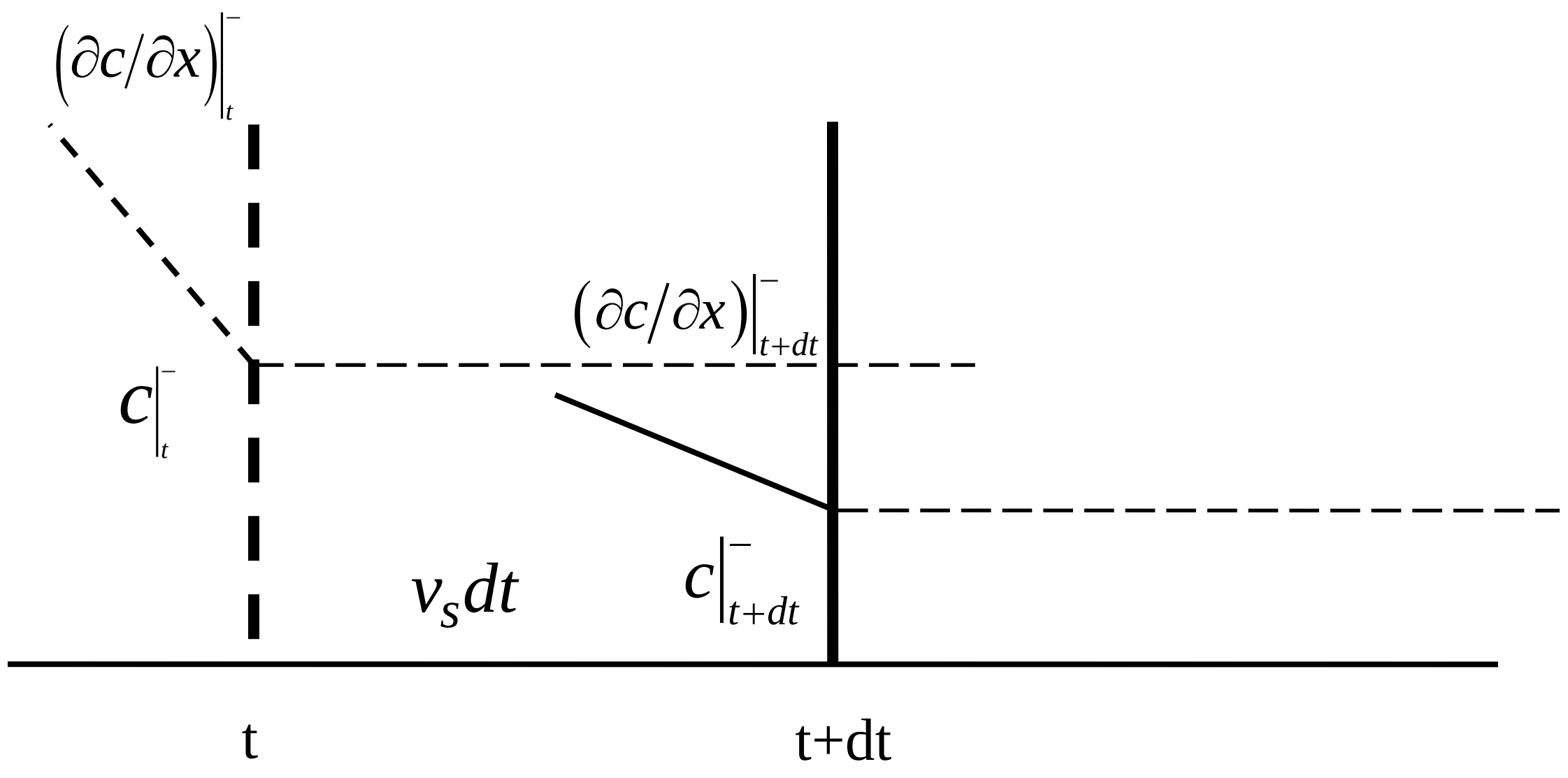

Fig.3.

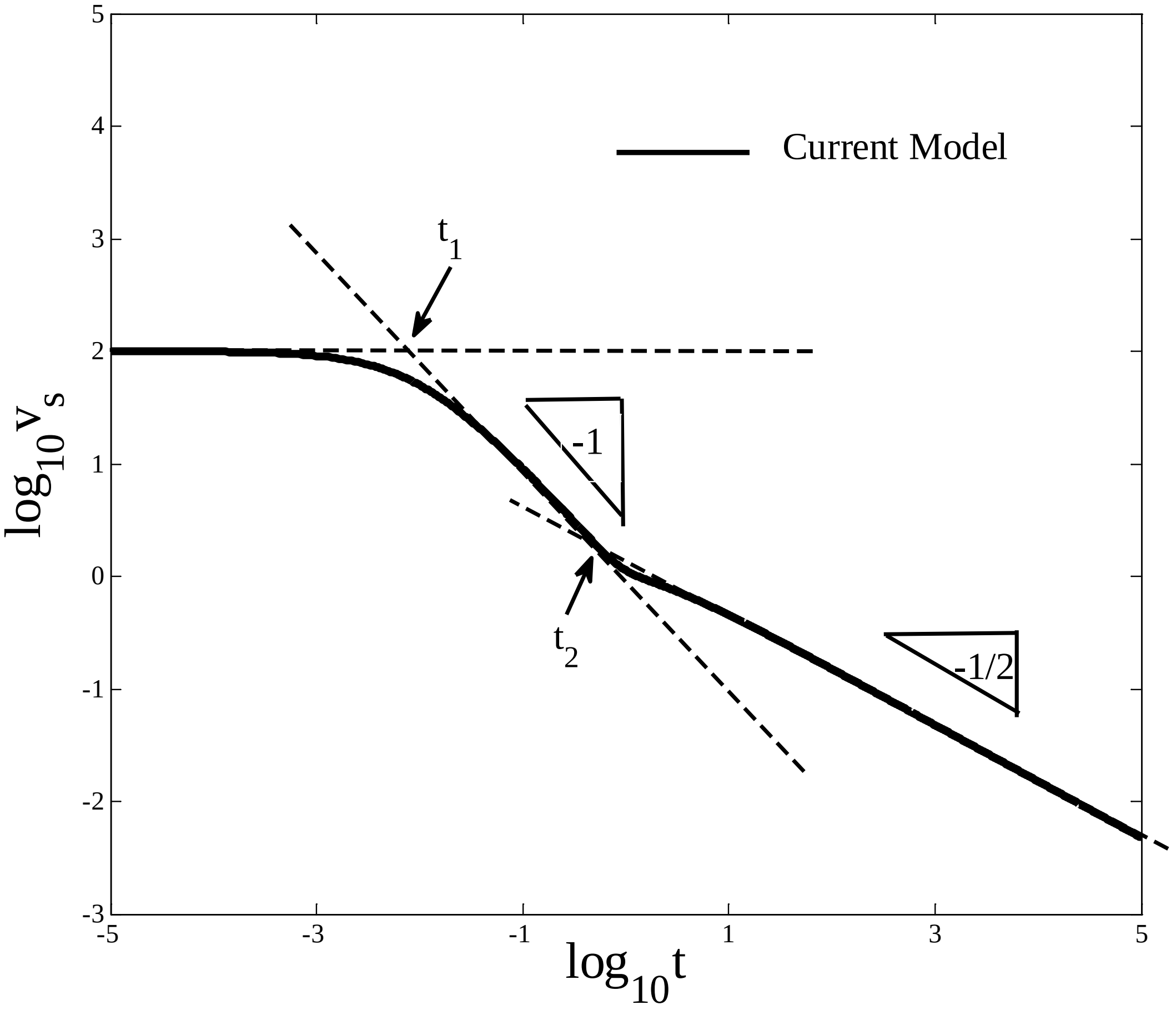

Fig.4.

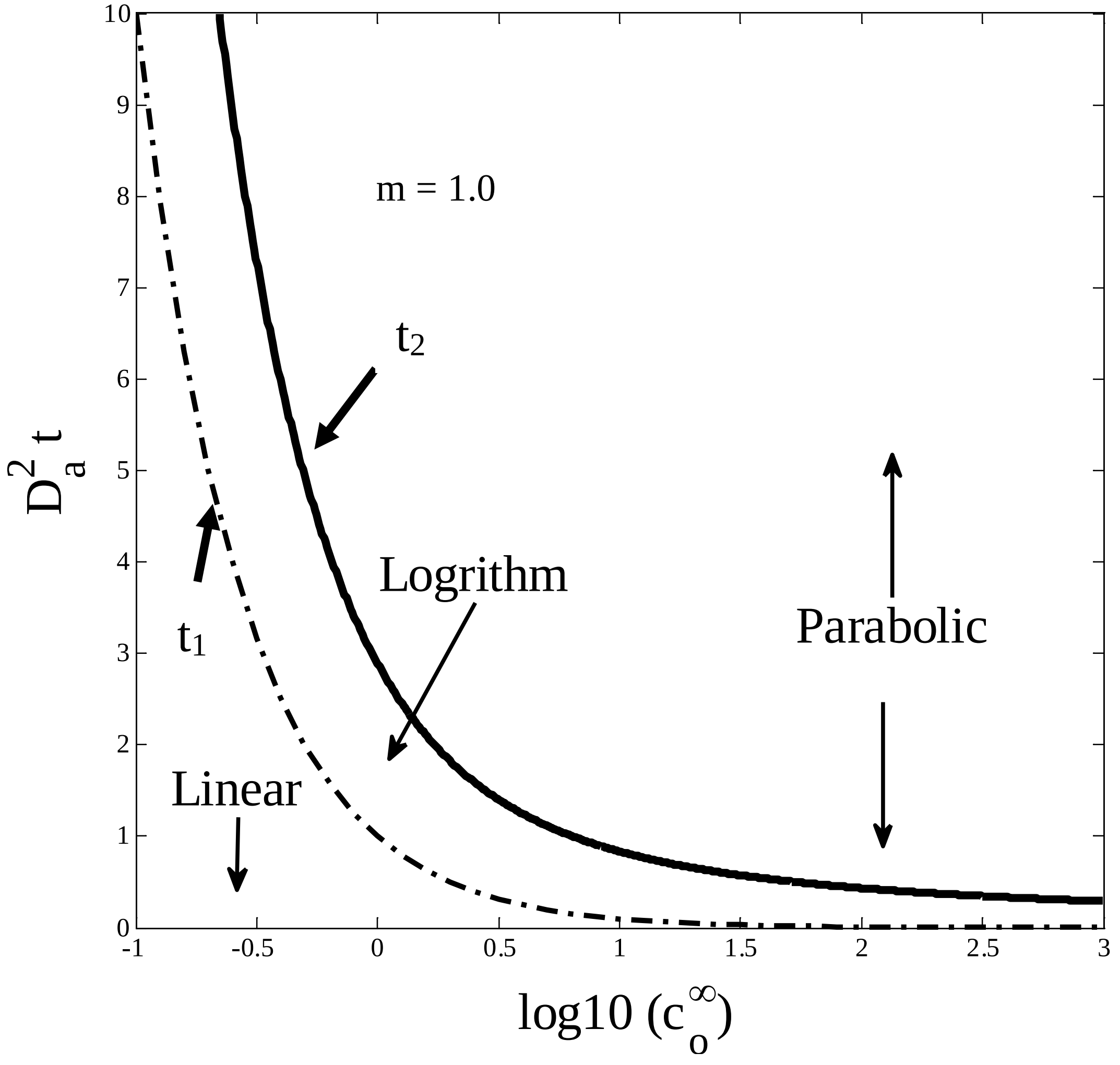

Fig.5.

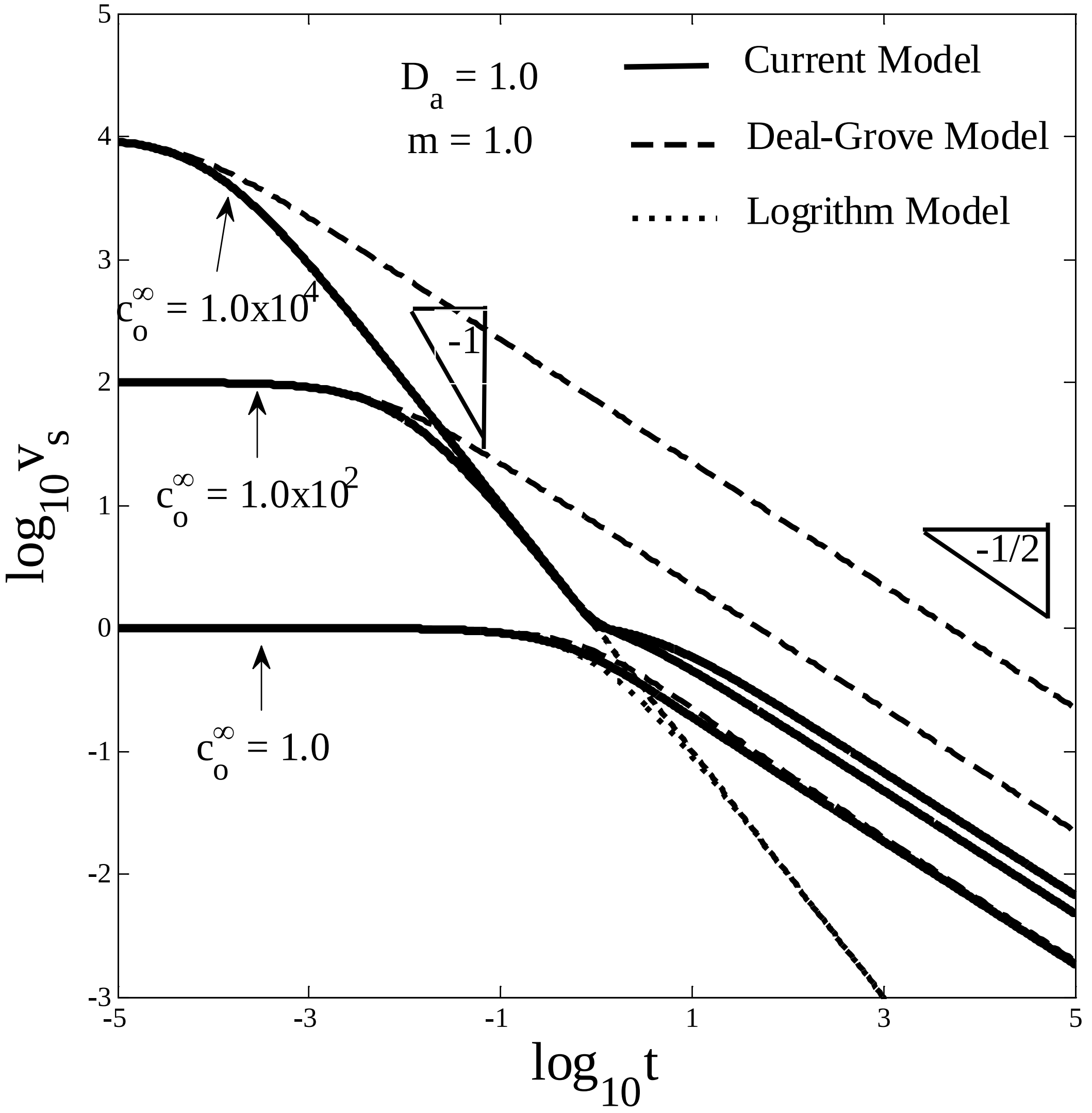

Fig.6.

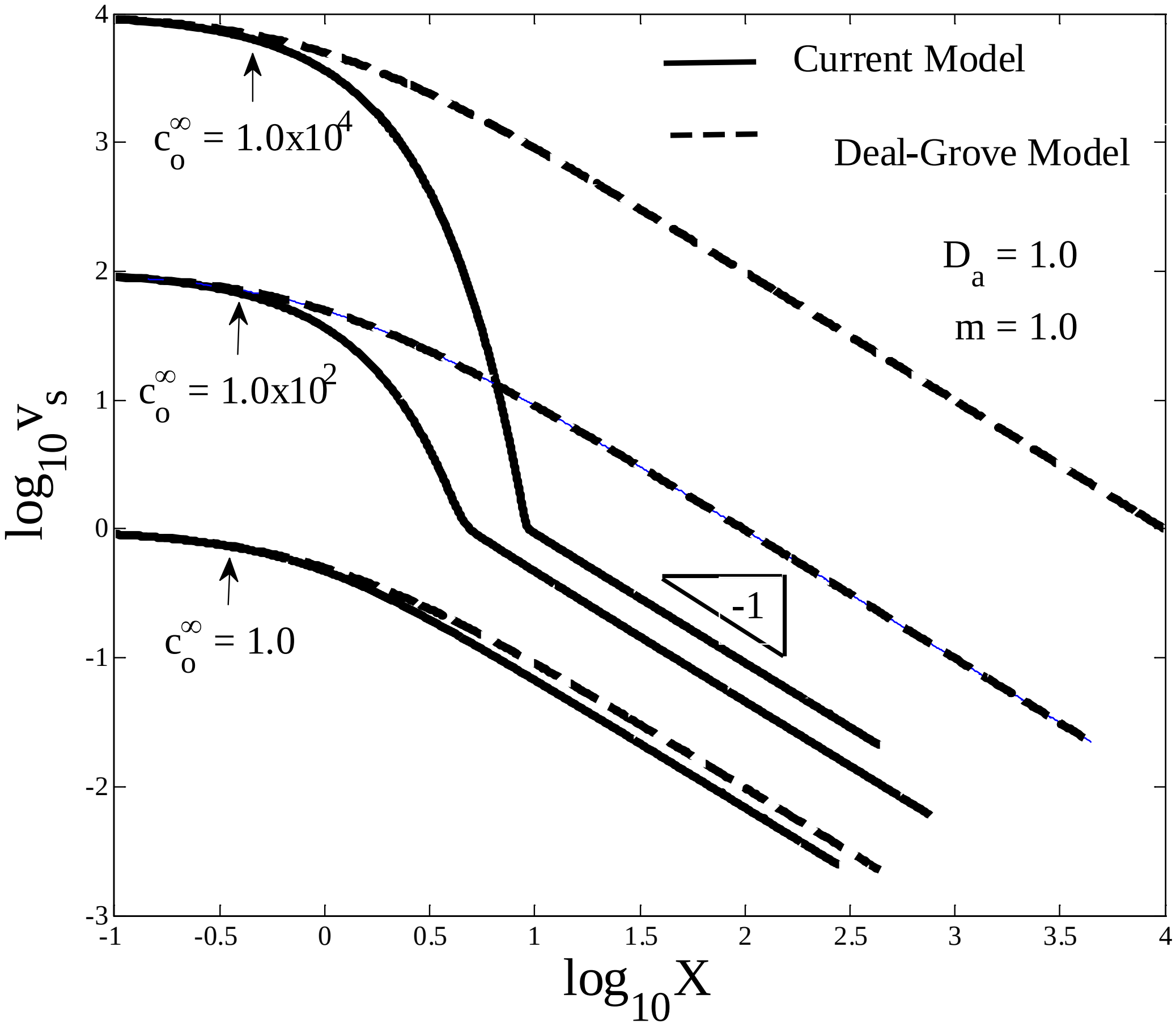

Fig.7.

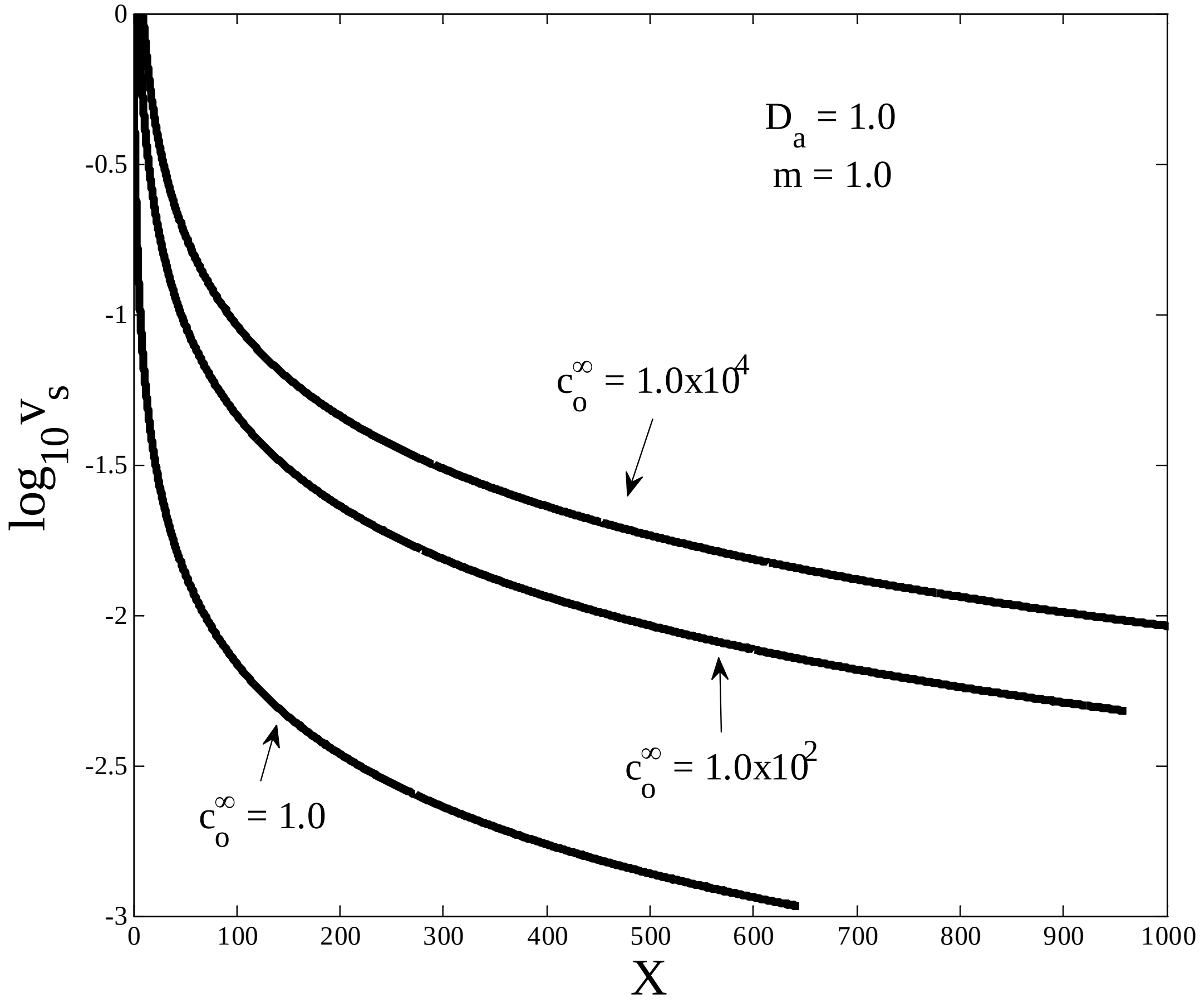

$D_a = 1.0$
$m = 1.0$
$c_o^\infty = 1.0x10^4$
$c_o^\infty = 1.0x10^2$
$c_o^\infty = 1.0$
$\log_{10} v_s$
X
0
-0.5
-1
-1.5
-2
-2.5
-3
0 100 200 300 400 500 600 700 800 900 1000

Fig.8.

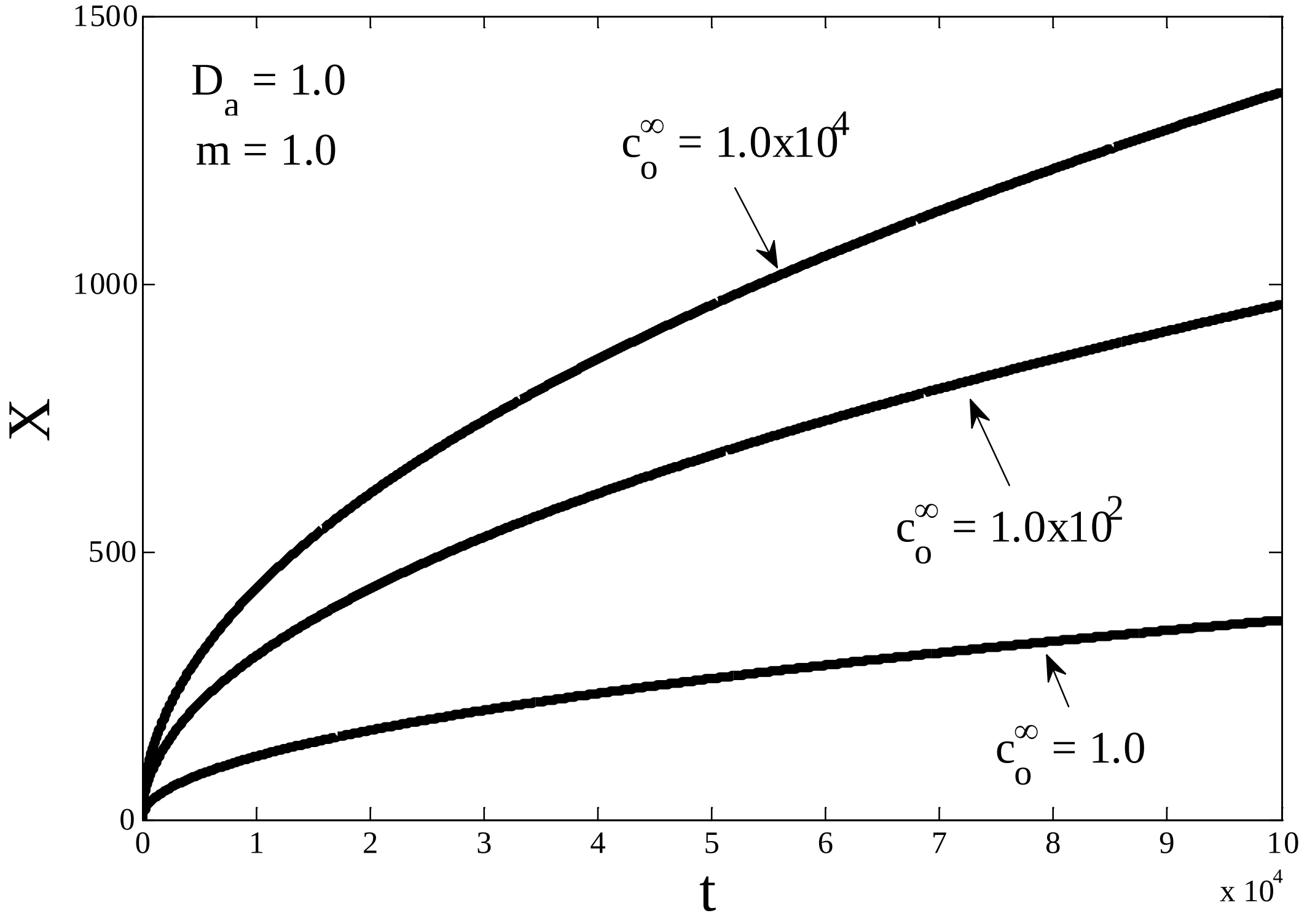

Fig.9.

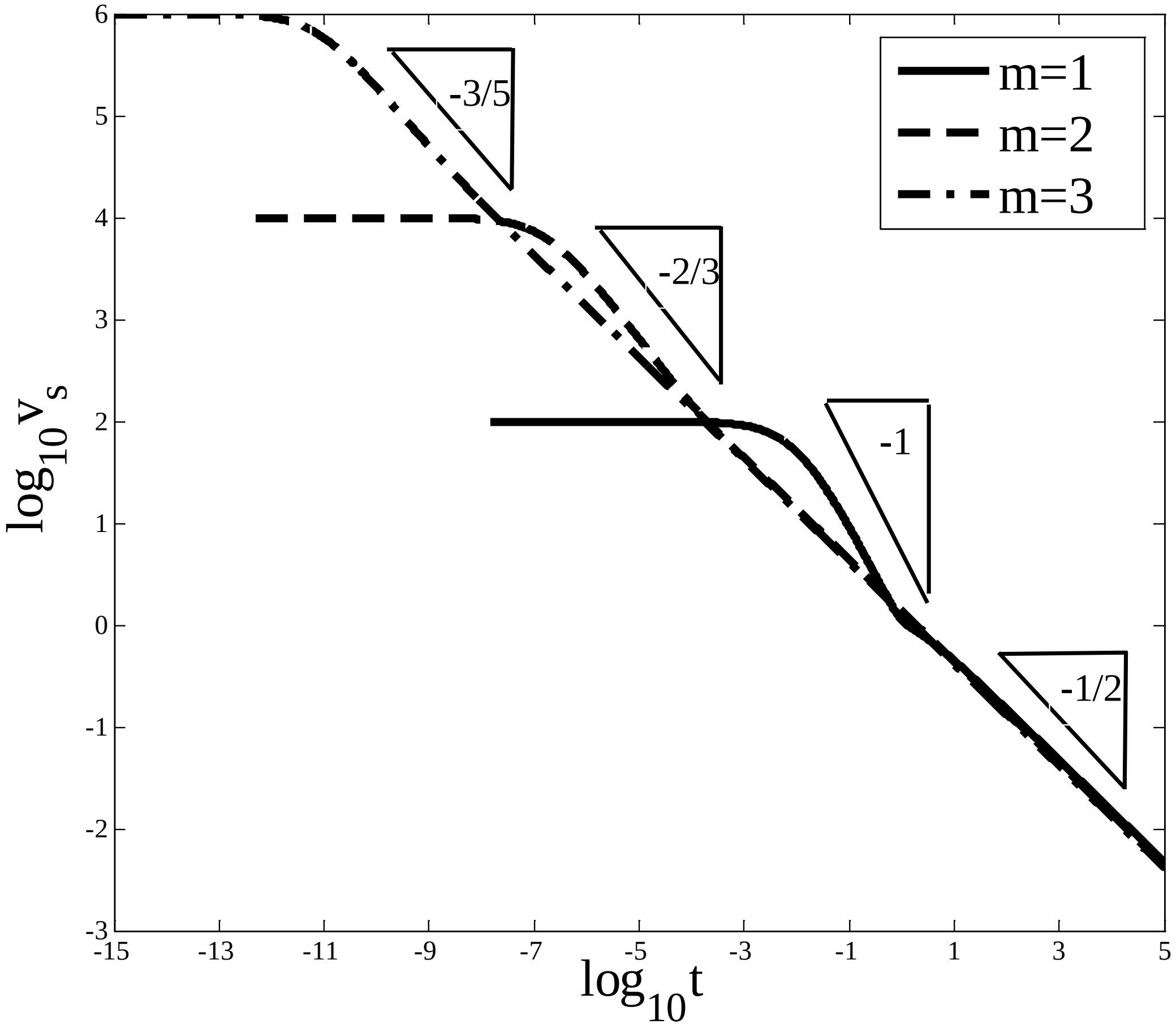